\documentclass[aps,twocolumn,showpacs,notitlepage,floatfix,prb,superscriptaddress,citeautoscript]{revtex4-1}

\usepackage[utf8]{inputenc}
\usepackage[dvipsnames,svgnames,table]{xcolor}
\usepackage{color}
\usepackage{colortbl}
\usepackage{amsmath}
\usepackage{soul}
\usepackage{multirow}
\usepackage[version=4]{mhchem}

\usepackage[english]{babel}
\usepackage[utf8]{inputenc}
\usepackage{amsmath}
\usepackage{graphicx,wrapfig,lipsum}

\usepackage{orcidlink}

\usepackage{hyperref}
\hypersetup{
    colorlinks=true,
    linkcolor=blue,
    filecolor=blue,      
    urlcolor=blue,
    citecolor=blue,
}
\usepackage{mathrsfs}
\usepackage{times}

\usepackage{bm}
\usepackage[T1,OT1]{fontenc}
\usepackage[normalem]{ulem}
\usepackage{fancyhdr}
\usepackage{float}

\hypersetup{
pdfstartview={FitH},
colorlinks=true,    
linkcolor=NavyBlue, 
citecolor=Maroon,   
filecolor=NavyBlue, 
urlcolor=NavyBlue   
}

\begin{document}

\title{
Symmetry-based computational search for novel binary and ternary 2D materials}

\newcommand{\jena}{Institut f\"ur Festk\"orpertheorie und -optik,
  Friedrich-Schiller-Universit\"at Jena and European Theoretical Spectroscopy Facility,
  Max-Wien-Platz 1, 07743 Jena, Germany}

\newcommand{\halle}{Institut f\"ur Physik, Martin-Luther-Universit\"at
  Halle-Wittenberg, D-06099 Halle, Germany}
  
\newcommand{\virginia}{Department of Physics, West Virginia University, Morgantown, WV 26506, USA}

\newcommand{\lux}{Department of Physics and Materials Science, University of Luxembourg, 162a avenue de la Fa\"iencerie, L-1511 Luxembourg, Luxembourg}

\author{Hai-Chen Wang\,\orcidlink{0000-0002-2892-5879}}
\affiliation{\halle} 
\author{Jonathan Schmidt\,\orcidlink{0000-0001-5685-6404}}
\affiliation{\halle} 
\author{Miguel A. L. Marques\,\orcidlink{000-0003-0170-8222}} 
\email{miguel.marques@physik.uni-halle.de}
\affiliation{\halle}
\author{Ludger Wirtz\,\orcidlink{0000-0001-5618-3465}}
\affiliation{\lux}
\author{Aldo H. Romero\,\orcidlink{0000-0001-5968-0571}}
\email{Aldo.Romero@mail.wvu.edu}
\affiliation{\virginia}
\affiliation{\lux}

\begin{abstract}
We present a symmetry-based {systematic} approach to explore the structural and compositional richness of two-dimensional materials. We use a ``combinatorial engine'' that constructs {candidate} compounds by occupying all possible Wyckoff positions for a certain space group with combinations of chemical elements. These combinations are restricted by imposing charge neutrality and the Pauling test for electronegativities. The structures are then pre-optimized with a specially crafted universal neural-network force-field, before a final step of geometry optimization using density-functional theory is performed. In this way we unveil an unprecedented variety of two-dimensional materials, covering the whole periodic table in more than 30 different stoichiometries of form A$_n$B$_m$ or A$_n$B$_m$C$_k$. Among the discovered structures, we find examples that can be built by decorating nearly all Platonic and Archimedean tesselations as
well as their dual Laves or Catalan tilings. We also obtain a rich, and unexpected, polymorphism for some specific compounds. We further accelerate the exploration of the chemical space of two-dimensional materials by employing machine-learning-accelerated prototype search, based on the structural types discovered in the {systematic} search. In total, we obtain around 6500 compounds, not present in previous available databases of 2D materials, with {a distance to the convex hull} of thermodynamic stability {smaller than 250~meV/atom}.
\end{abstract}

\maketitle

\section{Introduction}

Since the synthesis of single graphene layers~\cite{novoselov2004electric}, two-dimensional (2D) materials have attracted a significant interest from the community. Their relevance extends to different research fields, such as catalysis, electronic transport, optical properties, and topological properties. However, {the chemical space for 2D materials is still relatively unexplored}, even though great effort has been spent on investigating the vast chemical space for bulk, three-dimensional (3D) compounds. {In fact}, experimental synthesis efforts have focused on a few structures, mostly obtained by exfoliation of known 3D layered materials~\cite{Novoselov2005Jul}.

On the computational side, we can find a few online databases of 2D materials, such as Materials Cloud (MC2D)~\cite{mounet2018two}, V2DB~\cite{sorkun2020artificial}, 2DMatpedia~\cite{zhou20192dmatpedia}, and the Computational 2D Materials Database (C2DB)~\cite{haastrup2018computational,gjerding2021recent, Lyngby2022-us}. These databases were built starting from 3D databases, by exfoliating single-layers from layered, van der Waals compounds. At the moment the vast majority of known 2D materials correspond to binaries~\cite{sorkun2020artificial,song2021computational}. An exception is the very recent addition of materials discovered via a crystal diffusion variational autoencoder in Ref.~\onlinecite{Lyngby2022-us}.

These 2D databases are newer, and considerably smaller, than their three-dimensional counterparts, e.g. Materials Project~\cite{jain2013commentary}, the crystallographic open database (COD)~\cite{gravzulis2009crystallography}, the Cambridge structural database~\cite{allen2002cambridge}, the NIST crystallographic database~\cite{van2002nist}, OQMD~\cite{saal2013materials}, AFLOW~\cite{curtarolo2012aflow}, materials cloud~\cite{talirz2020materials}, and many others.
All these initiatives were seeded by experimental crystal structures stored in the Inorganic Crystal Structure Database (ICSD) and other experimental databases. In fact, the creation of the ICSD in 1912~\cite{bergerhoff1987crystallographic,zagorac2019recent, belsky2002new} paved the way to the {systematic} study of the relationship between crystal structure and materials properties. To complement experimental data, many databases (both 2D and 3D) also contain results from high-throughput studies (often accelerated by machine learning).

High-throughput searches are responsible for a majority of the calculations in the large theoretical databases like AFLOW~\cite{curtarolo2012aflow}, OQMD~\cite{saal2013materials} and DCGAT\cite{Schmidt_2022_MC}. Traditional high-throughput searches rely on simple empirical rules to select candidate materials for evaluation with density functional theory (DFT). Consequently, {they contain} a large number of highly unstable systems. 
A particularly popular approach is prototype search, where new materials are hypothesized by {changing} the chemical elements in a known crystal structure (often stemming from ICSD). In some cases, all combinations of chemical elements are taken into account, while in other cases arguments based on charge neutrality, atomic or ionic radii, etc. are used to circumvent the combinatorial nature of the problem.

{In comparison to these rule-based selections, machine learning algorithms generally allow us to consider all combinations of the chemical elements due to their computational efficiency~\cite{icgcnn, Schmidt_2022, chen2022universal, goodall2021rapid}. In fact, recent progress has enabled us to speed up the scanning of crystal prototypes by a factor of up to $\sim$2000~\cite{Schmidt_2022} with respect to traditional DFT high-throughput studies. A second research direction are generative models that do not rely on existing prototypes. Here, generative adversarial networks~\cite{zhao2021high,long2021constrained}, variational auto encoders~\cite{NOH20191370,ren2022invertible} and, more recently, diffusion  models~\cite{xie2021crystal,Lyngby2022-us} are the most successful approaches. While these generative models have made great progress over the last year and improved with respect to their bias towards stable structures, the stability of the structure still has to be evaluated with a secondary machine learning model.
No matter the generation or selection algorithm, the next step consists in a local structural optimization of each compound, invariantly using DFT as the workhorse method}~\cite{hohenberg1964inhomogeneous, kohn1965self}. Analysis of thermodynamic stability can then be achieved by computing the formation energy or the distance to the convex hull. In this way, databases have grown considerably and can now sometimes reach millions of crystal structures.

While the success of using chemical combinatorics is recognized {for 3D materials}, it has been a substantial handicap for predicting new 2D materials. The number of known 2D materials prototypes is unfortunately very small. Various research groups have considered different strategies to address this issue, often resorting to machine learning methods. This paper presents an entirely different approach which is not based on motifs or chemical substitutions. Instead, we create all possible combinations {of chemical elements} for binary (and ternary) systems for specific two-dimensional space groups. Therefore, based on symmetries and chemical criteria, we can arrive at a sizeable two-dimensional crystal structure database and a diverse set of structural prototypes. The crystal shapes show a variety of bondings and forms absent in existing databases. As our search is {systematic}, crystal structures contain a large number of different chemical formulae as well as almost all possible Wyckoff positions allowed by the space groups. Here we focus on two-dimensional materials but our approach is general, and it can also be applied to three-, one- or even zero-dimensional structures.

This article is structured as follows. We start by discussing in detail the {systematic} approach to discover 2D crystal structures and our strategies to accelerate the search. We then present an overview of the materials we discover, giving a few examples of structural diversity and polymorphism. In the following, we discuss machine-learning accelerated prototype search based on the wealth of prototypes obtained. In the Appendix following the conclusions, we give details on our methodology.

\section{Strategy}

\begin{figure}
\centering
\includegraphics[scale=0.33]{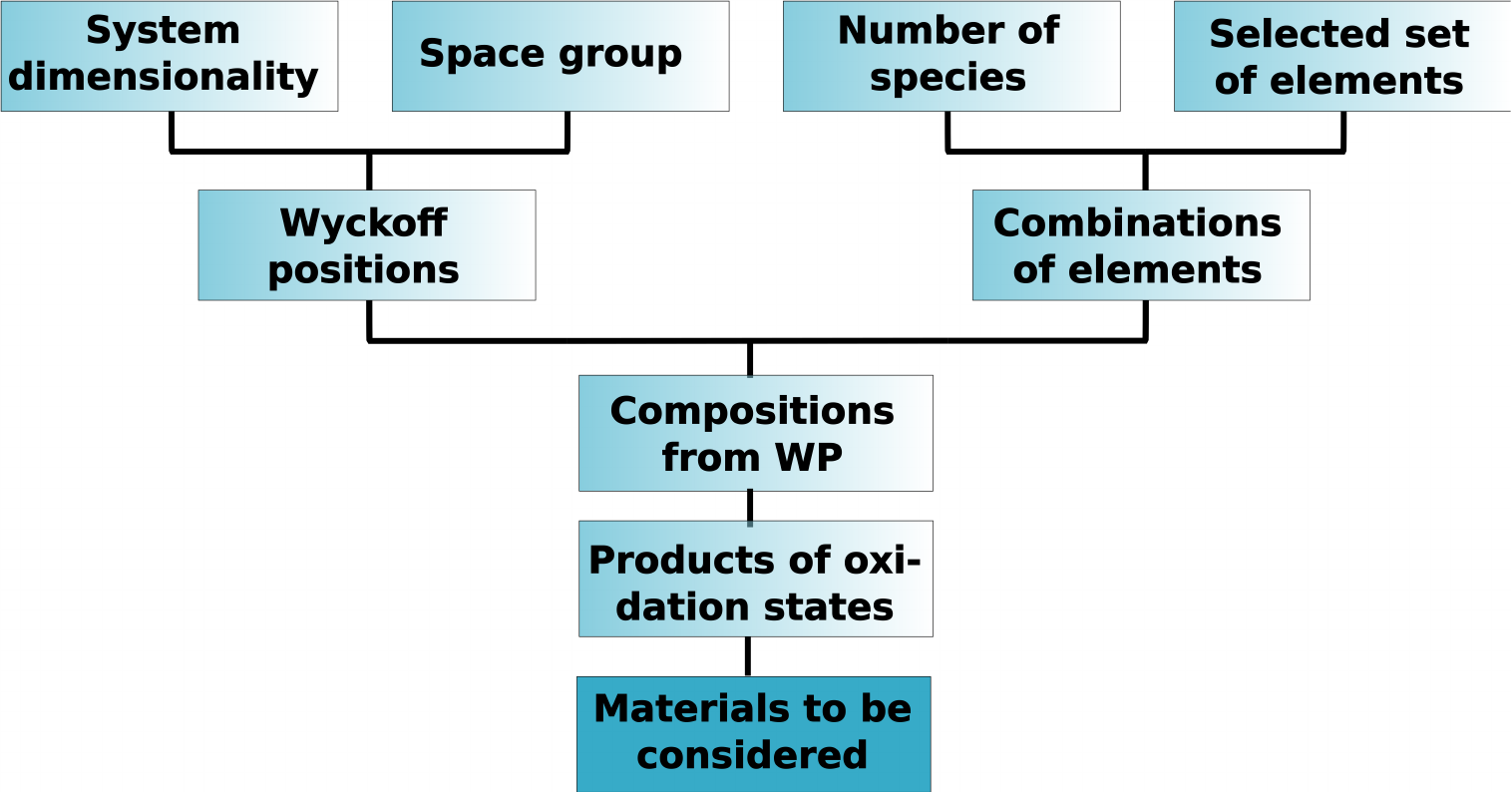}
\caption{Summarized flowchart for choosing potential materials using group symmetries and their Wyckoff positions.}
 \label{fig:flowchart}
\end{figure}

\begin{figure}
\centering
\includegraphics[scale=0.04]{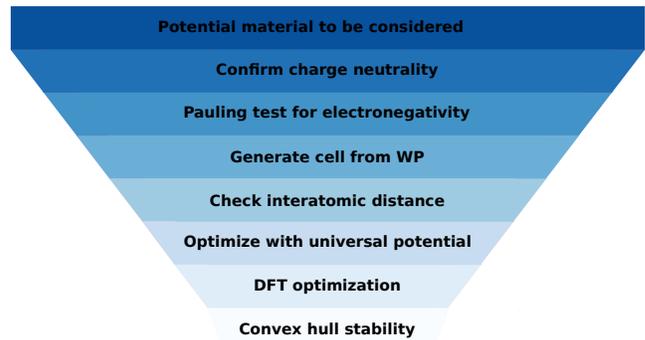}
\caption{Pipeline for the materials screening, from materials chosen following flowchart~\ref{fig:flowchart}.}
\label{fig:pipeline}
\end{figure}

Figure~\ref{fig:flowchart} and Fig.~\ref{fig:pipeline} summarize our approach for the generation of materials. The first step corresponds to a combinatorial workflow that creates hypothetical compounds. The initial input parameter is the number of desired chemical species in the particular material. We consider most elements of the periodic table, from Li to Bi, including first-row rare earth elements. We exclude, however, radioactive elements and rare gases, namely At, Tc, Pr, Pm, He, Ne, Ar, Kr, Xe, Rn. We then generate all possible combinations of the different elements selected from the periodic table. For example, a total of 2701 combinations are obtained with no repeated elements for a binary compound. 

The second parameter is the two-dimensional space group. We use the table of layer group symmetries, created by considering the wallpaper group and adding reflections in the perpendicular direction. A full description of the possible groups is given in Refs.~\onlinecite{litvin2002international, aroyo2011crystallography, aroyo2006bilbao}. To generate the atomic positions provided by the layer space group for all possible Wyckoff positions (WPs), we used the package PyxTal~\cite{fredericks2021pyxtal}. A list of the possible WPs can be found in the Bilbao Crystallographic Server. {From the 80 layered groups, we studied the 18 that have the smallest number of different Wyckoff positions and therefore the smallest number of combinations.}

The next step is the creation of all possible combinations of the WPs for each chemical element in the combined list. For example, if the number of WPs is four, we get $4!$ posibilities. This strategy allows different WPs for the same species and therefore broadens the number of possible stoichiometries (i.e., a chemical species can occupy more than one WP). We then create a product of this list associated with the number of selected species. This selection leads to the definition of a chemical composition based on the occupation of the different WPs for each species in the compound. {As certain Wyckoff positions have free parameters, our approach is not exhaustive. For example, for the $p1$ space group we only occupy the (single) position $1a$ once for each atomic species. This leads to a single possibility for both the binary compound AB, and  the ternary compound ABC. The number of possibilities increases, however, very rapidly with the number of different Wyckoff positions available within the space group.}

In parallel, we create a list of possible oxidation states of the considered species. We make all possible combinations without replacement for each element from this list, and we create a product of the different list elements. We used the experimentally most common oxidation states, as they will have considerably larger potential to be synthesized (the selected oxidation states are included in the Supplementary Information). Finally, a compound is created from the provided number of species, the combination of WPs, and the oxidation state.

\begin{table}[hth]
\caption{Crystallographic summary of the layer groups considered in this work: the space group symbol (and number in parenthesis), the Wyckoff positions (and site symmetries in parenthesis). We also show the number of binary ($N^{(2)}_\text{tot}$) and ternary systems ($N^{(3)}_\text{tot}$) generated by our combinatorial engine and the number of entries that were found below 250~meV/atom from the convex hull of stability ($N^{(2)}_{<0.25}$ {and $N^{(3)}_{<0.25}$)}.}
\label{Table1:WP}
\begin{tabular}{rccccc}
Space Group & Wyckoff Positions & $N^{(2)}_\text{tot}$ & $N^{(2)}_{<0.25}$ & $N^{(3)}_\text{tot}$ & $N^{(3)}_{<0.25}$ \\
\hline \\[-3mm]
$p1$       (01) &                              1a (1) &   225 &   5 \\
$p$11m     (04) &                    2b (1), 1a (..m) &  1321 & 339 \\
$p$11a     (05) &                              2a (1) &   225 &  47 & 1944 & 307 \\
$p211$     (08) &             2c(1), 1b(2..), 1a(2..) &  6645 & 623 \\
$p2_111$   (09) &                              2a (1) &   225 &  40 & 1944 & 273 \\
$c211$     (10) &                    4b (1), 2a (2..) &  1321 & 153 \\
$p$b11     (12) &                              2a (1) &   225 &  15 & 1944 & 448 \\
$c$m11     (13) &                    4b (1), 2a (m..) &  1321 & 268 \\
$p2_1$/b11 (17) &            4c (1), 2b (-1), 2a (-1) &  3129 & 220 \\
$p2_12_12$ (21) &              4c(1), 2b(..2),2a(..2) &  6645 & 398 \\
$p$b$2_1$m (29) &                    4b (1), 2a (..m) &  1321 & 140 \\
$p$b2b     (30) &          4c (1), 2b (.2.), 2a (.2.) &  6645 & 228 \\
$p$m2a     (31) &          4c (1), 2b (m..), 2a (.2.) &  6645 & 478 \\
$p$m$2_1$n (32) &                    4b (1), 2a (m..) &  1321 & 148 \\
$p$b$2_1$a (33) &                              4a (1) &   225 &  19 & 1944 & 139 \\
$p$b2n     (34) &                    4b (1), 2a (.2.) &  1321 &  74 \\
$c$m2e     (36) &          8c (1), 4b (m..), 4a (.2.) &  6645 & 383 \\
\multirow{2}{*}{p31m (70)} & 6d (1), 3c (..m),  & \multirow{2}{*}{15728} & \multirow{2}{*}{783} \\
             & 2b (3..) 1a (3.m)
\end{tabular}
\end{table}

{We have not imposed any explicit limit on the number of atoms in the unit cell. However, the procedure we use to generate the compounds does lead to an \textit{implicit} constraint which, however, depends on the number and multiplicity of the Wyckoff positions for each space group. For the space groups studied here, the maximum number of atoms in the unit cell was 32, although the majority of the compounds has less than 16 atoms in the unit cell.}

After the material is obtained from the previous step, and before we perform a complete electronic structure calculation, we conduct a materials screening, which allows us to reduce the number of compounds to be fully considered. For the screening, we used rules implemented in {the open-source material-screening Python} package SMACT~\cite{davies2016computational}. In this package, decisions are made based on stoichiometry. The first rule is to have only neutral compounds, which can be easily computed from the stoichiometry and the oxidation states. {The second rule} is the so-called Pauling test for neutral materials which requires that positive ions {have lower} electronegativity than negative ions. 

After screening, we use the main properties of a given material, such as oxidation states, stoichiometry, and Wyckoff positions to generate the potential compounds. {In this step,} we use the PyxTal utility~\cite{fredericks2021pyxtal} to create a {2D} unit cell {with the given} number of species. When WPs have internal degrees of freedom, PyxTal tries to create a unit cell with the provided symmetry constraints. First, the cell directions are selected according to the space group. Then, the WPs are generated from the symmetry operations, and, if there are internal degrees of freedom, they are set randomly. Next, the cell parameters and the volume are determined, assuming that each atom has a radius equal to the covalent bond radius. Finally, a density is obtained with the cell volume and atomic masses, which is compared with a minimum density. If the cell density is smaller than a minimum density (set to 0.75 in scaled units), the package attempts first to define the atomic positions randomly (setting the number of attempts to 40), and, in case this fails, it tries to change the cell parameters ({up to} 10 times) and repeat the generation of the atomic position. If a cell cannot be defined, the structure generation is considered unsuccessful, and a new material is considered. A summary of the pipeline is represented in figure~\ref{fig:pipeline}. {W}e generate {systematically} two dimensional structures for the space groups shown in Table~\ref{Table1:WP}. In this table we also include the corresponding Wyckoff positions, the site symmetry and the number of different compounds generated for each space group.

The next step is the geometry optimization. Unfortunately, the initial structures are usually very far away from equilibrium, making structural optimization with DFT cumbersome. To increase the efficiency of our workflow we perform an intermediate geometry optimization step using a universal neural-network force-field~\cite{chen2022universal}. {In contrast to standard force fields that are usually trained to reproduce the potential energy surface of a specific system, universal neural network force fields describe all possible compounds. Of course, the objective of the latter is not to replace the former, that will be more precise but with a more limited applicability. Instead, they provide a reasonable description for all geometrical arrangements and chemical elements. Our model, trained using a transfer learning approach, has a median absolute error of 96~meV/atom for geometry optimizations. This is already a competitive value, suitable for describing 2D materials in this intermediate screening step.}

At this point we remove from our dataset materials that are too thick (using a threshold of 7.5~\AA) {or that are} predicted to be too unstable by the machine learning model (more than 600~meV/atom from the hull, corresponding approximately to twice the MAE of the original model). {We also remove structures} that were already included in C2DB~\cite{gjerding2021recent} (excluding the very recent structures of Ref.~\onlinecite{Lyngby2022-us}).

The use of machine learning force fields resolves several technical problems: The pre-converged geometries are, in most cases, already quite good, only requiring a few steps of geometry optimization using DFT. They also allow us to discard many repeated and very high-energy structures. After the DFT geometry optimization we evaluate the distance to the convex hull of stability. We use the convex hull of Ref.~\onlinecite{Schmidt_2022_MC,Schmidt_2022} that is considerably larger than the one of the Materials Project~\cite{jain2013commentary}, in particular in what concerns the ternary (and quaternary) sector. Consequently, our distances to the hull are sometimes larger than in other 2D databases.

{Note that besides thermodynamic stability, the issue of dynamical stability is a crucial factor for 2D materials, and should always be verified before a specific material is proposed for synthesis. A material is dynamically stable when it exhibits no imaginary phonon frequencies across the Brillouin zone. Unfortunately, the calculation of the phonon dispersion is extremely time-consuming, and even more so for 2D systems due to issues related to the vacuum required to treat the long-range part of the Coulomb interaction.~\cite{PhysRevB.96.075448} We also note that imaginary phonon frequencies could be an indication of a charge-density wave phase (at even lower formation energy) which we might be overlooking due to the use of unit cells with a limited number of atoms.}

\section{2D Materials}

\begin{figure}[htb]
\centering
\includegraphics[scale=0.25]{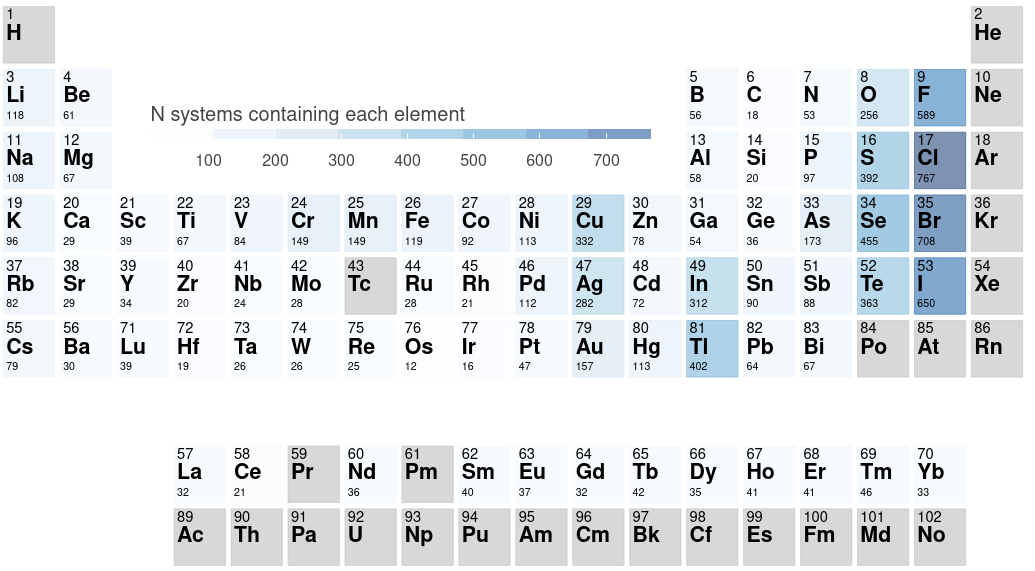}

(a)

\includegraphics[scale=0.25]{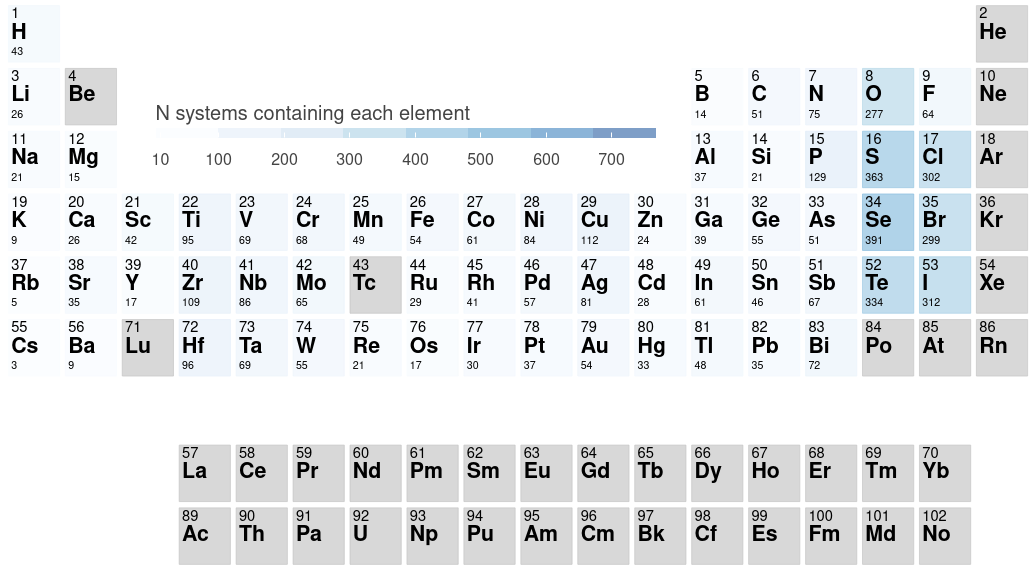}

(b)
\caption{Periodic tables showing the frequency of the chemical elements in binary compounds below 250~meV/atom in (a)~our work and (b)~C2DB. We emphasize that our entries do not include, by construction, any of the compounds of C2DB.}
 \label{fig:comparingdb}
\end{figure}

\begin{figure}[htb]
\centering
\includegraphics[scale=1.33]{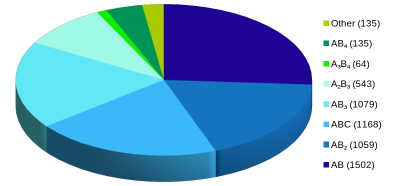}
\caption{Pie chart of the frequency of different general chemical formula for compounds with distances to the convex hull smaller than 250~meV/atom in our database. The category \textit{other} includes 29 general formulas.}
 \label{fig:Stoichio}
\end{figure}

It turns out that our workflow was able to arrive at the large majority of systems already present in C2DB. This is particularly true for binary systems, as these were more extensively investigated than ternaries (see Table~\ref{Table1:WP}). This, in our opinion, fully validates our workflow.

Figure~\ref{fig:comparingdb} presents a comparison of the binary materials present in our database (excluding the ones found in C2DB) compared to C2DB. For the discussion, we only took into account the materials that are within a distance of 250~meV/atom from the convex hull of stability, that corresponds loosely to the definition of ``high-stability'' in C2DB~\cite{gjerding2021recent}. Note that for consistency we have reoptimized the C2DB structures using our convergence criteria and our selected set of pseudopotentials.

\begin{figure}[htb]
\centering
\includegraphics[width=.98\columnwidth]{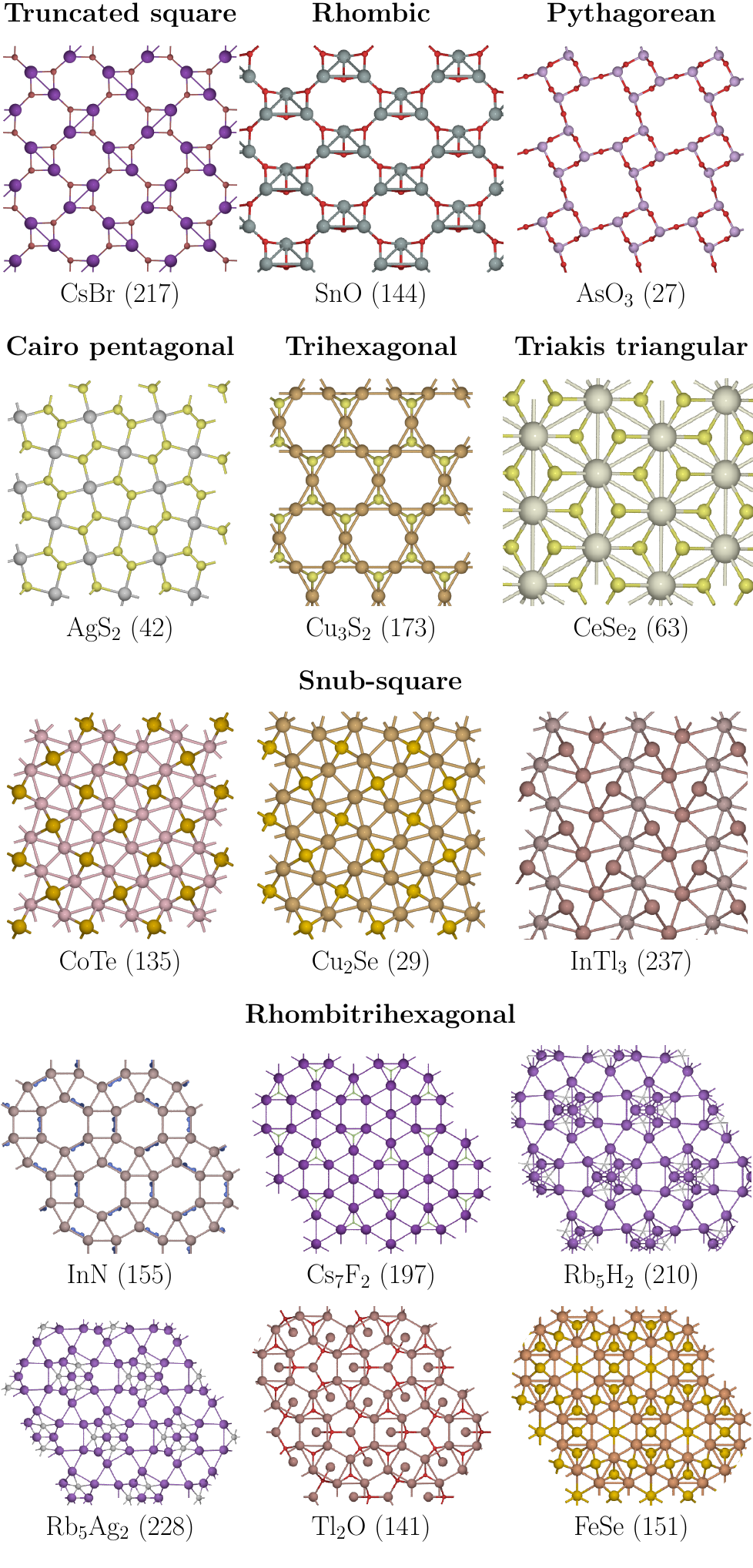}
\caption{Diversity of structures obtained by our procedure. We label in bold the convex tiling from which the structure can be derived. Next to the chemical formula, in parenthesis, is the distance to the convex hull (in meV/atom) of the specific compound. Note that the structure depicted is not necessarily the lowest 2D phase for the composition we found in our search.}
 \label{fig:diversity}
\end{figure}

We find 2D compounds across the whole periodic table, including some with lanthanides that have been up to now excluded from previous works. Not surprisingly, the majority of compounds includes a non-metal element (due to the requirement of charge neutrality), leading to the pronounced peaks for O, S, Se, Te, F, Cl, Br, I, etc. The figure also reveals some differences in the prevalence of certain elements between our dataset and C2DB. For example, we find considerably more compounds with F than with O, while in C2DB we observe the opposite behavior. As our approach is to a large extent {systematic} in what concerns chemical compositions and geometries, we believe that the differences are explained by a bias already present in ICSD and other databases that were used to seed the 2D databases. For example, it is well known that oxides are over-represented in experimental works as they can be more easily synthesized and are often stable in air.

Other conclusions can be drawn from Fig.~\ref{fig:comparingdb}. For example, it can be seen clearly that the {non-metals in the second row} have more difficulty in forming low-energy compounds than other non-metals in the same group. This is a consequence of the Singularity Principle~\cite{restrepo2004topological}, i.e., that the chemistry of the these elements is often different to the later members of their respective groups. Furthermore, elements like N, O, C, and F form very strong directional covalent bonds that leave {comparatively little room for distortions that would be required to form different structures}. {As for} metallic elements, {it is in particular} the transition elements {in the fourth row} from Ti to Cu (and in particularly this last one), together with late group III-A (In and Tl) seem to form easily 2D compounds.

The diversity of stoichiometries is illustrated in Fig.~\ref{fig:Stoichio}. As our emphasis {has been} on binary compounds, it is not surprising that most represented stoichiometries are binary. Among these, the simple \ce{AB2}, \ce{AB3}, \ce{A2B3}, etc. dominate the low-energy structures. This fact can be easily understood by the requirement of charge neutrality and the fact that most non-metals have oxidation states of -I, -II, or -III. As such, the same situation can be found for bulk, 3D semiconductors and insulators. However, we do find a long list of other stoichiometries (more than 30), and these often reveal very interesting and unexpected structures.

In Fig.~\ref{fig:diversity} we give a glimpse of the diversity of structural motifs found by our method. Note that this is far from a complete list of all 2D structures found. We concentrate on unusual arrangements that go beyond the most common square and hexagonal lattices. We emphasize that these motifs appeared naturally in our workflow and {were not} constructed by hand. Interestingly, we easily found examples that can be derived from the majority of the different Euclidean uniform tilings, both Platonic and Archimedean as well as their dual Laves or Catalan tilings. Moreover, many of these tilings seem to be unique to the two-dimensional world, as no layered 3D material is known to possess them.

The first two structures can be derived from a truncated square and a rhombic tiling. In the first case, \ce{Cs2Br2} squares are connected, forming regular empty octahedra, leading to a rather open lattice. In the second, \ce{Se3O3} rectangular units form bonds along the corners, leading to flattened octahedra. We then present an example of a Pythagorean tiling, a motif that is composed of two different squares that share one side, and can be found all over the world in kitchen or garage floors. Interestingly, it was proposed recently that elementary, two-dimensional Cl, Br, and I might be able to adopt this arrangement~\cite{huran2021atomically}. We also find a series of Cairo pentagonal tilings. In this example, \ce{AgS2} forms two overlaying tessellations of the plane by irregular hexagons, where each of the hexagons is formed by four identical pentagons. At the center of the hexagons we find a Se--Se bond. 
{Note that this is the same Cairo pentagonal tiling that was found for \ce{PdSe2}}~\cite{LV2021101231,10.1021/jacs.7b04865,C8TC06050A}.
One of the possible structures of \ce{Cu3S2} consists on a trihexagonal tiling (that is often called the Kagome
lattice due to its use in traditional Japanese basketry) of the plane by Cu atoms, decorated by a S atom in the middle of the triangles. Triakis triangular lattices appear quite commonly in our data. The example in Fig.~\ref{fig:diversity} can be seen as composed of Ce equilateral triangles decorated with a Se atom at its center.

The following three examples are derived from snub-square lattices. In the first two, the metal forms this interesting square-triangle lattice and the non-metal decorates the squares. In the case of \ce{CoTe}, Te-atoms can be found above and below the plane of the Co atoms, while in \ce{Cu2Se}, Se-atoms alternate above and below the plane of the Cu-atoms. We note that this specific lattice was recently proposed for some noble metal chalcogenide monolayers~\cite{gao2021semiconducting} and for certain Ba and Ti oxides~\cite{huran2021two}. The third example is more complex, as both In and Tl form a distorted version of the snub-square lattice, with further Tl-atoms alternating above and below the plane. Note, however, that this curious structure is almost at the limit of our energy threshold. Finally, we present six examples of rhombitrihexagonal lattices, where the metal atoms form the triangle-square-hexagon lattice that is then decorated (mostly) by the non-metals. We found a very diverse number of different decorations, allowing for many different stoichiometries, ranging from the simple \ce{AB} and \ce{AB2} to the more unconventional \ce{A2B5} and \ce{A2B7}. A very interesting possibility raised by the finding of all these snub-square and rhombitrihexagonal structures is that these can be easily inflated by a recursive approach to generate quasi-crystalline systems~\cite{forster2013,forster2020}. This is, however, only possible for structures not including out-of-plane alternating atoms, as this induces frustration in the system reducing its stability~\cite{huran2021two}.

\begin{figure}[htb]
\centering
\includegraphics[width=7.5cm]{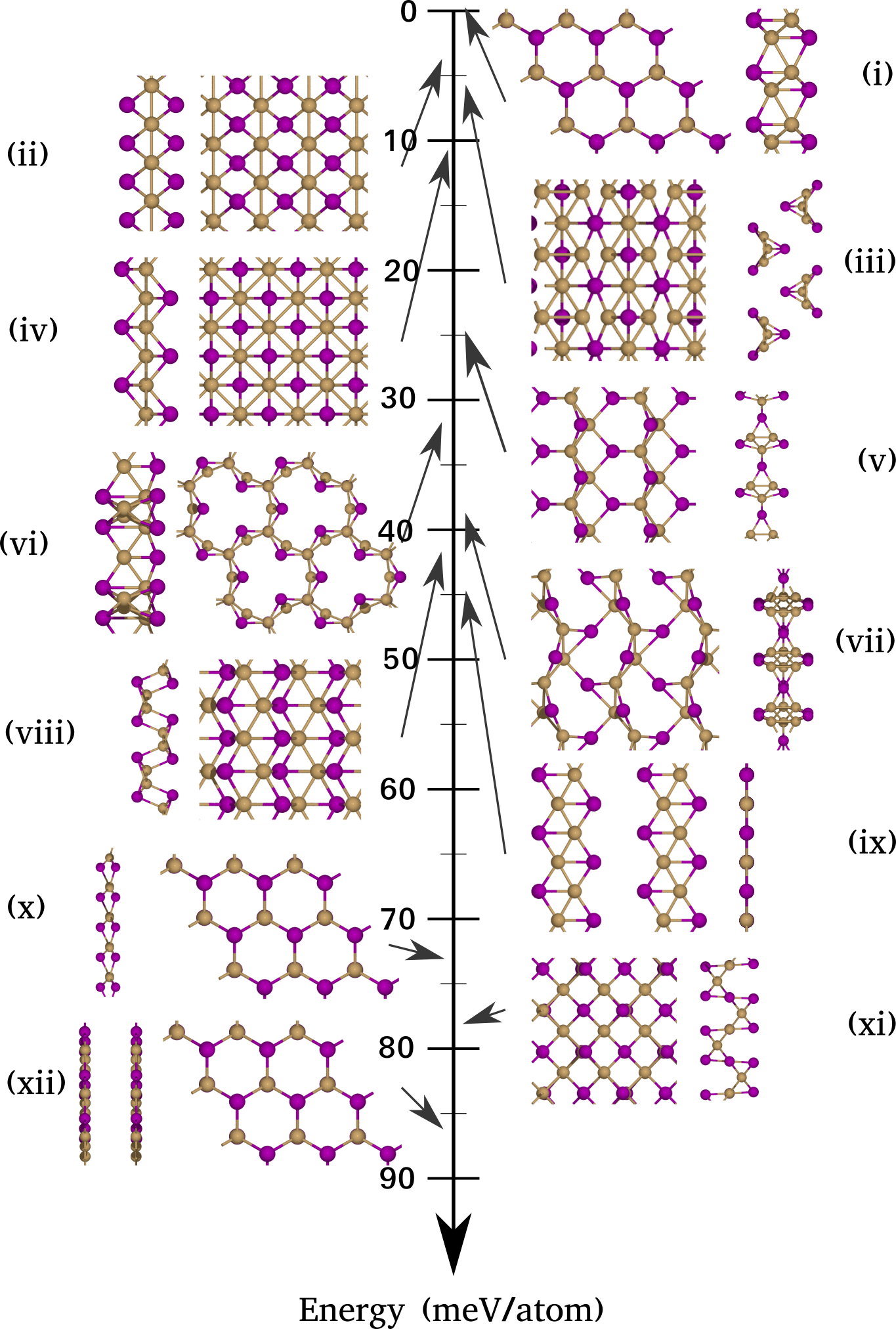}
\caption{Polymorphism in CuI. The vertical axis denotes increasing energy distance to the convex hull of stability in meV/atom. Copper atoms are depicted in gold, while iodine is in violet. For each structure we show a top and a side view.}
 \label{fig:CuI}
\end{figure}

We have given several examples of the different stoichiometries in our dataset and of the structural variety stemming from the{m}. Now, we look at the issue of polymorphism, i.e., the different phases possible for a specific chemical composition. Not surprisingly, polymorphism depends strongly on the chemical elements present in the compound. For example, for BN we found a single structure below 250~meV/atom, the well-known honeycomb lattice, while for other compounds we have an extraordinary variety in the same energy range.

As an example, we show in Fig.~\ref{fig:CuI} a selection of the crystal structures {that} we found for CuI. We recall that zincblende CuI is at the moment the most promising $p$-type transparent conducting semiconductor~\cite{grundmann2013cuprous}. However, CuI has a number of polymorphs, including a a couple of trigonal phases~\cite{kurdyumova1961cui,sakuma1988crystal,villars2009cui,akopyan2010specific} that are layered, with a bonding pattern rather different from the $\gamma$-phase. Fig.~\ref{fig:CuI} shows that also in the 2D case, we find a large variety of structures and of bonding patterns.

As the lowest-energy 2D layer we find a covalently bound hexagonal double-layer (i) that is essentially on the convex hull of thermodynamic stability. The (buckled) single layer (x) and the van-der-Waals bound double flat-layer also appear in the energy spectrum but considerably higher, at more than 70~meV/atom. The second most stable structure is, surprisingly, a rectangular lattice of Cu--I (ii), with the I-atoms alternating above and below the plane of the Cu-atoms. A related lattice (iv) appears just a few meV/atom above. Structure (iii), {which} is only 6~meV/atom above the hull, and {structure} (ix) are arrangements of one-dimensional objects. The first exhibits nanowires with a triangular section arranged in an alternating fashion as depicted in Fig.~\ref{fig:CuI}. The latter (ix) is a periodic arrangements of nanostripes. (Incidentally, higher in energy, at 161~meV/atom, we even find a molecular crystal of \ce{Cu4I4} pyramidal clusters). All these systems turn out to be semiconducting, with calculated (PBE) band gaps ranging from around 0.5~eV to more than 2.1~eV.

\section{Prototype search}

The {biggest} advantage of the workflow presented above is that it is (i)~{systematic} and (ii)~unbiased in what concerns the structural variety. Unfortunately, the price to pay for these advantages is efficiency, in the sense that it is computationally expensive to go through all possible compositions and space groups and that many of the possibilities turn out to be highly unstable or lead to thick slabs. It is, however, possible to accelerate considerably the exploration of the 2D material space by using the structural prototypes discovered by our approach, and combining them with a machine-learning model appropriate for prototype search~\cite{Schmidt_2019,Kulik_2022}. Of course, in this way we will not discover new structural motifs, but we can explore the whole compositional space very efficiently.

Our approach follows the same basic principles as V2DB~\cite{sorkun2020artificial}, but goes beyond it in a number of different directions. First, we perform transfer learning from a 3D machine, which allows to transfer many of the chemical principles that govern atomic bonding. Second, we use a much larger training set, increasing the accuracy of the machine. Third, we lift several constrains (like charge neutrality or electronegativity rules) used in V2DB and in our {systematic} search, and we expand the possible chemical elements to the whole periodic table. This allows us to discover a variety of intermetallics and compounds combining elements with unusual oxidation states. We furthermore perform machine-learning predictions for all two-dimensional prototypes, either already present in C2DB or stemming from our {systematic} search. Finally, we perform validation DFT calculations for some of the predictions, specifically for the binary stoichiometries \ce{A2B5}, \ce{A2B7}, and the ternary \ce{ABC2}, \ce{ABC3}, \ce{AB2C2}, \ce{A2B2C3}.

{Note that, in contrast to the systematic generation of structures based on the space groups, in the machine-learning assisted prototype search, we do not impose any constraint on the possible oxidation states. As such, the machine can, and does, propose 2D systems including chemical elements in other, less common oxidation states.}

To keep the number of structures manageable, we asked the machine to output all structures that it found below 200~meV/atom for the binaries and 50~meV/atom for the ternaries. In total, we obtained 1023 candidates that were pre-optimized with our neural-network force-field and then optimized with DFT. From these 638 were found to be below 250~meV/atom from the hull, yielding an exceptional success rate of around 62\%. The lowest success rate, of only 9\%, was found for the \ce{A2B7} stoichiometry: as these compounds were sparsely present in the training set, the machine could not learn the specificity of those structures. The problem can, of course, be solved by adding further samples to the dataset, in order to remove the structural (and compositional) bias, as previously shown for bulk systems in Ref.~\onlinecite{Schmidt_2022}.  We are currently performing DFT calculations for $\sim 40000$ more materials, resulting from  238 million machine-learning predictions, that will be available in the next release of our dataset.

\section{Conclusions}

We have presented a {systematic} approach to explore the structural and compositional diversity that is possible in the chemical space of 2D materials. The main advantage of this approach is that it is not based on a specific number of structural prototypes. This is particularly important for 2D materials, as the space of possible structures is still rather unexplored and only few prototype structures, mainly from exfoliation of layered 3D materials, are known so far. In this way, we have discovered thousands of unexpected phases that have no counterpart in the world of layered three-dimensional materials. We expect that such unusual bonding patterns and geometrical patterns will also lead to unique mechanical, electronic, optical, and magnetic properties.

Our method relies heavily on the use of machine learning. {The} extensive use of neural networks in several parts of our workflow is self-accelerating. In fact, the faster we generate more data for two-dimensional systems, the larger will be our training sets, resulting in even more accurate machine learning models. This leads to a virtuous cycle that, in our opinion, will pave the way for a rather complete exploration of the binary, ternary, and eventually also the quaternary two-dimensional phases in the near future.

Finally, an important question is how many of the phases in our dataset can be synthesized. We chose to filter our results to include only compounds with an energy less than 250~meV/atom above the convex hull, as these have higher stability and therefore higher probability to be synthesized. (For comparison, silicene, that has been experimentally synthesized\cite{Vogt2012Apr}, is more than 600~meV/atom above the hull in its free-standing form.) However, besides these thermodynamic arguments, a key factor will be the choice of suitable substrates that stabilize the two-dimensional layers, and, of course, the ingenuity of experimental physicists {and chemists} to design targeted synthesis strategies.

\section{Appendix: Methods}
\label{sec:methods}

\subsection{DFT calculations}

We performed all geometry optimizations and total energy calculations with the code \textsc{vasp}~\cite{vasp1,vasp2}.  The 2D Brillouin zones were sampled by uniform $\Gamma$-centered k-point grids with a density of 6~$k$-points/\AA$^{-2}$. We performed spin-polarized calculations starting from a ferromagnetic state, and used the projector augmented wave setups~\cite{paw,paw2} of \textsc{vasp} version 5.2 with a cutoff of 520~eV. We converged the calculations to forces smaller than 0.005 eV/\AA. As exchange-correlation functional we used the Perdew-Burke-Ernzerhof~\cite{PBE} functional with on-site corrections for oxides {and} fluorides containing Co, Cr, Fe, Mn, Mo, Ni, V, {or} W. The repulsive on-site corrections to the d-states were
3.32, 3.7, 5.3, 3.9, 4.38, 6.2, 3.25, and 6.2~eV, respectively. These parameters were chosen to be compatible with the Materials Project database~\cite{jain2013commentary}. We imposed a vacuum region of 15~\AA, and systems that resulted in structures with a thickness greater than 7.5~\AA\ were automatically discarded. Finally, as it is common in this kind of approaches, some of the calculations did not converge due to a multitude of reasons. The corresponding phases were then simply eliminated from the dataset.

Distances to the convex hull were evaluated using \textsc{pymatgen}~\cite{pymatgen2013CompMatSci} using the large complex hull of Ref.~\onlinecite{Schmidt_2022} corresponding to the dataset available in the Materials Cloud repository~\cite{Schmidt_2022_MC}.

\subsection{\textsc{m3gnet}}

We employed the universal neural-network force-field \textsc{m3gnet}~\cite{chen2022universal} that was developed to reproduce the energies and the forces of bulk structures with remarkable results. As a starting point we used the pretrained network distributed with \textsc{m3gnet}. We tested this model on 1300 of our systems by measuring the difference between the energy calculated with \textsc{m3gnet} (at the \textsc{m3gnet} relaxed structure) and the energy calculated with DFT (at the DFT relaxed structure). We arrived at a mean absolute error (MAE) of 320~meV/atom and a median absolute error of 223~meV/atom. These numbers are already rather small, especially when we consider that the training set of \textsc{m3gnet} did not include 2D systems that can exhibit very different bonding patterns {compared to bulk structures}. {As soon as} enough data was available from our own simulations, we used transfer learning techniques to retrain \textsc{m3gnet} for 2D materials (see Section~\ref{sec:methods}). 
Specifically, we build a dataset comprising energies, forces, and stresses from structures calculated during the geometry optimization steps. Structures with extremely high forces above 50~eV/\AA\ were removed from the data as were structures with no neighboring atoms inside the cutoff radius to avoid errors during training. To balance the training set, for systems with more than 4 recorded geometry optimization steps only the first, last and N$_\text{steps}$/3 step were used.
The final training set for \textsc{m3gnet} contained 11612 geometry relaxations corresponding to 34944 energies and structures. 
The resulting network had a validation MAE of 61~meV/atom for direct energy predictions after training. The test errors for geometry optimizations on the same dataset as the pretrained model were 198~meV/atom for the MAE and 96~meV/atom for the median absolute error proving the efficiency of our transfer learning strategy. Of course, we expect these errors to decrease further simply by adding more data to the training set. The models were trained with the base hyperparameters from \textsc{m3gnet} and by setting the loss function of the stress in the non-periodic direction to zero.

\subsection{Crystal-graph attention networks}

We used the crystal-graph attention neural networks developed in Ref.~\onlinecite{Schmidt_2021} as they were specifically crafted for prototype searches. In particular, they require as input only the (unrelaxed) structural prototype and not accurate relaxed structures. Of course, this model was trained on bulk 3D structures, so we do not expect it to perform accurately in our case. However, many of the bonding patterns present in our 2D materials can already be found in the 3D world. To take advantage of this, we performed transfer learning of the 3D model, using the 2D structures in our dataset as training data.
We used a dataset of DFT calculations with 22007 entries, 80\% of which were used for training 10\% for validation and 10\% for testing. Evaluating both models on the test set, we arrive at an MAE of 222~meV/atom for the original model and 86~meV/atom for the model transferred to the 2D data.

\begin{figure}[htb]
    \centering
    \includegraphics[scale=0.5]{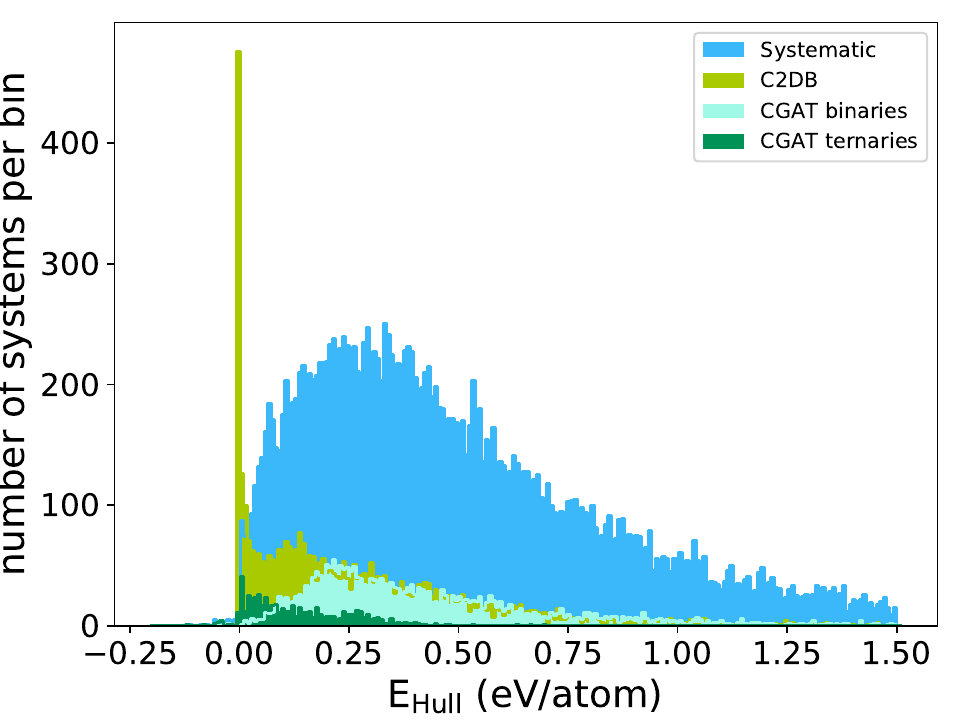}
    \caption{Distribution of the distances to the convex hull (calculated with DFT) for the compounds obtained through our {systematic} approach and stemming from the machine-learning assisted prototype search compared to the entries of C2DB.}
    \label{fig:ehull}
\end{figure}

In Fig.~\ref{fig:ehull} we present a histogram with the distances to the convex hull of stability calculated with DFT. The C2DB data is in light green, and is highly peaked at zero, decaying slowly for larger energies. This is expected, as C2DB was seeded with stable 3D, van der Waals bonded structures from ICSD. In blue we depict the structures obtained through our {systematic} approach. These form a continuous distribution with a peak at around 300~meV/atom, and extending beyond 1.5~eV. Knowing that such distribution for random compounds can extend beyond 4~eV, we see how the charge and electronegativity constraints lead to relatively stable compounds (at the price of overlooking intermetallics or compounds with unusual oxidation states. In light blue we show the machine-learning binaries predicted to be within 200~meV/atom from the hull. This shows a peak at around that value, as expected form the cutoff, then decaying similarly to the C2DB data. The ternary entries, displayed in green, are shifted to much lower energy, as expected by the smaller cutoff of 50~meV/atom. This results are absolutely consistent with the MAE of 86~meV/atom for the 2D model.
The CGAT-hyperparameters are listed in the supplementary material and the code can be found at \url{https://github.com/hyllios/CGAT.git}.

\section{Data availability statement}
All relevant data used in or resulting from this work is available at Materials Cloud (\url{https://doi.org/10.24435/materialscloud:sb-cy}). The database, including structures, distances to the hull, and other basic properties, can be accessed at \url{https://tddft.org/bmg/physics/2D/} through a simple web-based interface.

\section{Acknowledgements}

We acknowledge the computational resources awarded by XSEDE, a project supported by National Science Foundation grant number ACI-1053575. The authors also acknowledge the support from the Texas Advances Computer Center (with the Stampede2 and Bridges supercomputers). We also acknowledge the Super Computing System (Thorny Flat) at WVU, which is funded in part by the  National Science Foundation (NSF) Major Research Instrumentation Program (MRI) Award \#1726534, and West Virginia University. 
AHR and LW were funded in part, by the Luxembourg National Research Fund (FNR), Inter Mobility 2DOPMA, Grant Reference 15627293. For the purpose of open access, and in fulfillment of the obligations arising from the grant agreement, the authors have applied a Creative Commons Attribution 4.0 International (CC BY 4.0) license to any Author Accepted Manuscript version arising from this submission. AHR also recognizes the support of West Virginia Research under the call research challenge grand program. JS and MALM gratefully acknowledge the Gauss Centre for Supercomputing e.V. (www.gauss-centre.eu) for funding this project by providing computing time on the GCS Supercomputer SUPERMUC-NG at Leibniz Supercomputing Centre (\url{https://www.lrz.de}) under the project pn25co.
JS and MALM gratefully acknowledge the computing time provided to them on the high-performance computers Noctua 2 at the NHR Center PC2. These are funded by the Federal Ministry of Education and Research and the state governments participating on the basis of the resolutions of the GWK for the national highperformance computing at universities (www.nhr-verein.de/unsere-partner).

We also thank Kristian Thygesen for kindly providing us with full access to the C2DB database. 

\section{Author  Contributions}
AHR performed the systematic structure generation; MALM and HCW performed the DFT high-throughput calculations; JS performed the training of the machines and the machine learning predictions of the distance to the hull; AHR, LW, and MALM directed the research; all authors contributed to the analysis of the results and to the writing of the manuscript.

\section{Competing  Interests}
The authors declare that they have no competing interests.

\bibliography{references.bib}

\end{document}